\documentclass[a4paper]{PoS}
\usepackage[a4paper,top=5.9cm,bottom=0.7cm,left=5.5cm,right=0.4cm]{geometry}
\usepackage{float}
\usepackage{amsmath}
\usepackage{subcaption}
\usepackage{booktabs}
\usepackage[utf8]{inputenc}
\usepackage[labelfont=bf,skip=0.5em,font=small]{caption}
\usepackage{multicol}

\makeatletter
\renewenvironment{thebibliography}[1]
   {\begin{multicols}{2}[\section*{\refname}]%
     \@mkboth{\MakeUppercase\refname}{\MakeUppercase\refname}%
      \list{\@biblabel{\@arabic\c@enumiv}}%
           {\settowidth\labelwidth{\@biblabel{#1}}%
            \leftmargin\labelwidth
            \advance\leftmargin\labelsep
            \@openbib@code
            \usecounter{enumiv}%
            \let\p@enumiv\@empty
            \renewcommand\theenumiv{\@arabic\c@enumiv}}%
      \sloppy
      \clubpenalty4000
      \@clubpenalty \clubpenalty
      \widowpenalty4000%
      \sfcode`\.\@m}
     {\def\@noitemerr
       {\@latex@warning{Empty `thebibliography' environment}}%
      \endlist\end{multicols}}
\makeatother

\title{Non-perturbative renormalization of tensor bilinears in Schr\"odinger Functional schemes\footnote{
IFT-UAM/CSIC-15-10, FTUAM-15-41}}

\ShortTitle{Non-perturbative renormalization of tensor bilinears in SF schemes}

\author{Patrick Fritzsch,$^{a}$ Carlos Pena,$^{a,b}$ \speaker{David Preti}$^{a}$\\
        \llap{$^a$} Instituto de F\'isica Te\'orica UAM/CSIC, Universidad Aut\'onoma de Madrid,\\
        C/\,Nicol\'as Cabrera 13-15, Cantoblanco, Madrid 28049\\
        \llap{$^{b}$} Departamento de F\'isica Te\'orica, Universidad Aut\'onoma de Madrid,\\
        Cantoblanco, Madrid 28049\\
        E-mail: \email{p.fritzsch@csic.es}, \email{carlos.pena@uam.es}, \email{david.preti@csic.es}}
 
\abstract{We present preliminary result for the study of the renormalization group evolution of tensor bilinears in Schr\"odinger Functional (SF) schemes for $N_f=0$ and $N_f=2$ QCD with non-perturbatively $\mathcal{O}(a)$-improved Wilson fermions. First $N_f=2+1$ results (proceeding in parallel with the ongoing computation of the running quark masses \cite{Campos:2015fka}) are also discussed. A one-loop perturbative calculation of the discretisation effects for the relevant step scaling functions has been carried out for both Wilson and $\mathcal{O}(a)$-improved actions and for a large number of lattice resolutions. We also calculate the two-loop anomalous dimension in SF schemes for tensor currents through a scheme matching procedure with RI and $\overline{\rm MS}$.  Thanks to the SF iterative procedure the non-perturbative running over two orders of magnitude in energy scales, as well as the corresponding Renormalization Group Invariant operators, have been determined.}

\FullConference{The 33rd International Symposium on Lattice Field Theory\\
		14 -18 July 2015\\
		Kobe International Conference Center, Kobe, Japan}

\begin{document}

\section{Introduction}
Tensor currents play an important r\^ole in interesting  processes through which the consistency of the Standard Model (SM) is being currently probed, as e.g. rare meson decays (see e.g. \cite{Buchalla:2008jp,Antonelli:2009ws} and references therein) or precision measurements of $\beta$-decays and limits on the neutron electric dipole moment (see e.g.  \cite{Bhattacharya:2011qm}).
Considering the tensor operator for two generical (and formally distinct) flavours $\psi_1$,$\psi_2$ as $T_{\mu\nu}=i\overline{\psi}_1(x)\sigma_{\mu\nu}\psi_2(x),$
its $\mathcal{O}(a)$ improvement in the chiral limit is achieved by considering the combination
\begin{equation}
T_{\mu\nu}^I=T_{\mu\nu}+ac_T\left (\tilde{\partial}_{\mu}V_{\nu}-\tilde{\partial}_{\nu}V_{\mu} \right)
\end{equation}
where the coefficient $c_T$ was computed at 1-loop within the Schr\"odinger Functional (SF) in \cite{ Sint:1997dj} and reads $c_T=0.00896(1)C_Fg_0^2+\mathcal{O}(g_0^4)$, for $C_F=(N^2-1)/2N$ and N colours. 
At vanishing spatial momentum, the only non-vanishing two-point functions with boundary operators allowed by the SF boundary conditions involve the "electric" components (see Eq. \eqref{eq.ktcorr} below)
\begin{equation}
T_{0 k}^I=T_{0 k}+ac_T\left (\tilde{\partial}_{0}V_{k}-\tilde{\partial}_{k}V_{0} \right)
\end{equation}
where the second term in the parenthesis of the rhs vanishes when inserted in Eq. \eqref{eq.ktcorr}.
This operator renormalizes multiplicatively; i.e. the corresponding operator insertion in any on-shell renormalized correlation function is given by 
\begin{equation}
\bar{O}(x,\mu)=\lim_{a \to 0} Z(g_0,a\mu)O(x,g_0),
\end{equation}
where $g_0$, $a$ are the bare coupling and the lattice spacing respectively and $\mu$ is the renormalization scale. The renormalization group running is described by the Callan-Symanzik equations 
\begin{equation}
\mu\frac{\partial}{\partial\mu}\bar{O}(x,\mu)=\gamma(\bar{g}(\mu))\bar{O}(x,\mu),
\quad \quad \quad 
\mu\frac{\partial}{\partial \mu}\bar{g}(\mu)=\beta(\bar{g}(\mu)).
\label{eq.CS}
\end{equation}
Since we will work in a mass independent scheme (i.e. renormalization conditions will be imposed in the chiral limit),  the $\beta$-function and all anomalous dimensions will only depend on the renormalized coupling $\bar{g}(\mu)$ and they can be expanded perturbatively in powers of ${g}$ as
\begin{equation}
\beta(g) \stackrel{g \sim 0}{\approx} -g^3(b_0+b_1g^2+b_2g^4+\ldots),\, \quad \quad  \quad \gamma(g) \stackrel{g \sim 0}{\approx} -g^2(\gamma_0+\gamma_1g^2+\gamma_2g^4+\ldots)\,,
\end{equation}
with universal coefficients $b_0$, $b_1$, $\gamma_0$ given by
\begin{equation}
b_0=\frac{1}{(4\pi)^2}\left \{ 11-\frac{2}{3}N_f \right \}, \quad b_1=\frac{1}{(4\pi)^4}\left \{ 102-\frac{38}{3}N_f \right \}, \quad \gamma_0=\frac{2C_F}{(4\pi)^2}.
\end{equation}
All the other coefficients of the expansions are scheme dependent.
\section{Renormalization in SF schemes}
In order to impose renormalization conditions we introduce a SF two-point function of the tensor current with boundary sources of the form
\begin{equation}
k_T(x_0)=-\frac{a^6}{2}\sum_{\mathbf{y},\mathbf{z}}\langle T_{0k}(x_0)(\bar{\zeta}(\mathbf{y})\gamma_k\zeta(\mathbf{z}))\rangle
\label{eq.ktcorr}
\end{equation}
and the respective improved correlator is then $k_T^I(x_0)=k_T(x_0)+ac_T\tilde{\partial}_0k_V|_{x_0}.$
In order to avoid extra divergences arising from the boundary the correlator $k_T$ can be normalized with boundary-to-boundary correlators
\begin{equation}
k_1=-\frac{a^{12}}{6L^6}\sum_{\mathbf{y},\mathbf{z},\mathbf{y'},\mathbf{z'}}\langle(\bar{\zeta}'(\mathbf{y'})\gamma_k\zeta'(\mathbf{z'}))(\bar{\zeta}(\mathbf{y})\gamma_k\zeta(\mathbf{z}))\rangle
\end{equation}
\begin{equation}
f_1=-\frac{a^{12}}{6L^6}\sum_{\mathbf{y},\mathbf{z},\mathbf{y'},\mathbf{z'}}\langle(\bar{\zeta}'(\mathbf{y'})\gamma_5\zeta'(\mathbf{z'}))(\bar{\zeta}(\mathbf{y})\gamma_5\zeta(\mathbf{z}))\rangle.
\end{equation}
The (mass independent) renormalization conditions then read
\begin{equation}
Z_T^{(\alpha)}(g_0,L/a)\frac{k_T^{I}(L/2)}{f_1^{1/2-\alpha}k_1^{\alpha}}=\left . \frac{k_T(L/2)}{f_1^{1/2-\alpha}k_1^{\alpha}} \right |_{m_0=m_c,g_0=0}.
\label{eq.renorm}
\end{equation}
The freedom in the choice of $\alpha$ in \eqref{eq.renorm} and in the angle $\theta$ entering spatial boundary conditions \cite{Sint:1993un} define a class of renormalization schemes. In the present work we have considered $\alpha=0,1/2$ and $\theta=0,0.5,1$. Following the standard SF iterative renormalization procedure \cite{Capitani:1998mq}, we define step scaling functions (SSF) as 
\begin{equation}
\Sigma_T^{(\alpha)}(u,a/L)=\frac{Z_T^{(\alpha)}(g_0,a/2L)}{Z_T^{(\alpha)}(g_0,a/L)} 
\label{eq.ssf}
\end{equation}
The continuum SSF for a bilinear correlator \cite{Capitani:1998mq, DellaMorte:2005kg} and for the coupling \cite{Bode:2001jv,Luscher:1993gh} are defined respectively by 
\begin{equation}
\sigma_T(g^2)=\exp \left \{ \int_g^{\sqrt{\sigma(g^2)}} \, dg' \, \frac{\gamma(g')}{\beta(g')} \right \} 
\quad \rightarrow \quad  
\sigma_T(u)=\lim_{a \to 0} \Sigma_T(u,a/L),
\end{equation}
\begin{equation}
-\log(2)=\int_g^{\sqrt{\sigma(g^2)}} \, dg' \, \frac{1}{\beta(g')}
\quad \rightarrow \quad
\sigma(u)=\bar{g}^2(2L), \, u=\bar{g}^2(L),
\end{equation}
where the index $\alpha$ has been suppressed.
%\vspace{-0.01\textwidth}
\section{Perturbative one-loop computation}
\begin{figure}[t]
\begin{subfigure}{0.51\linewidth}
\includegraphics[width=\linewidth]{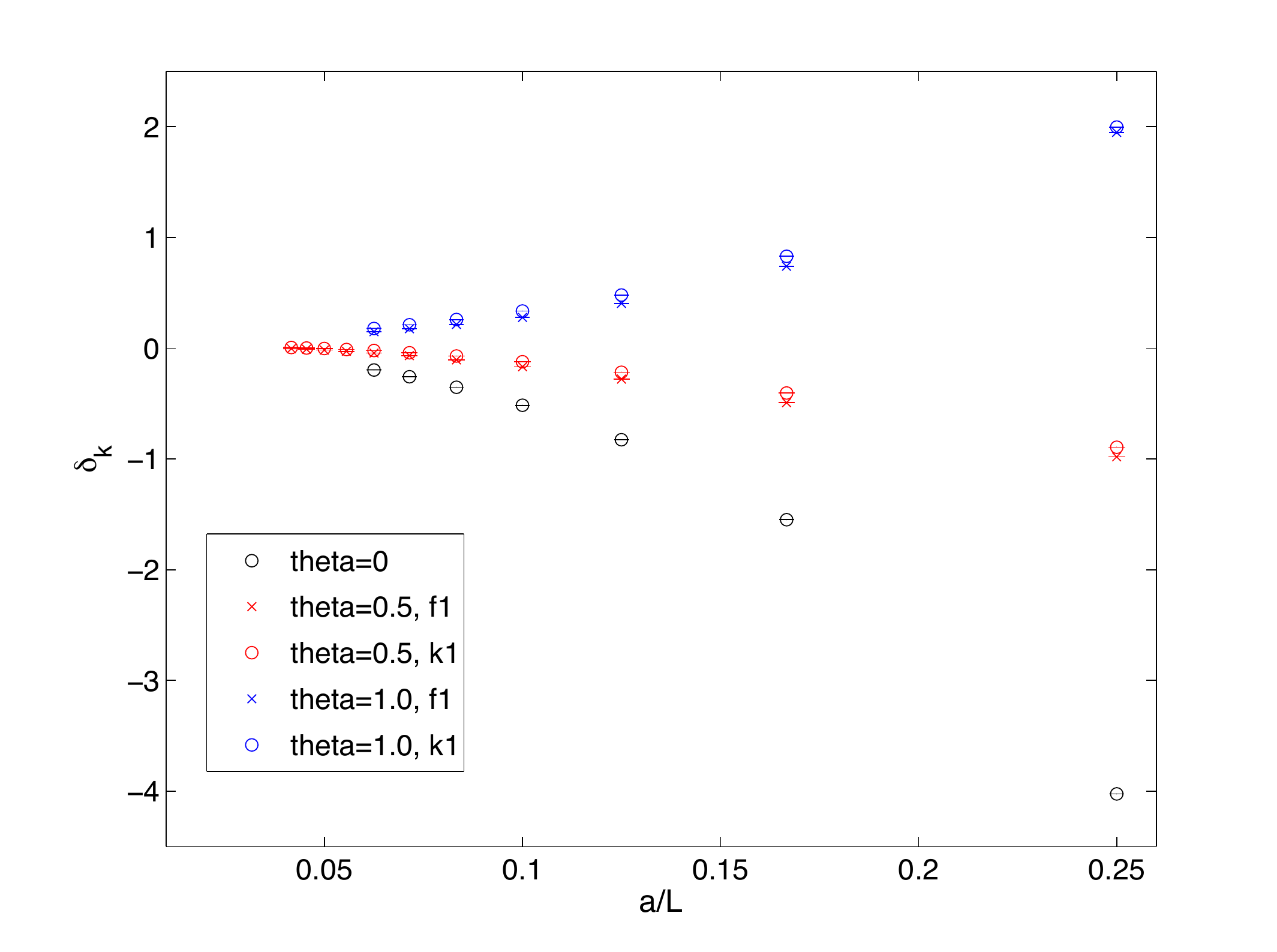}
\label{fig:cutoff_L}
\end{subfigure} \hspace{0.02\linewidth}
\begin{subfigure}{0.51\linewidth}
\includegraphics[width=\linewidth]{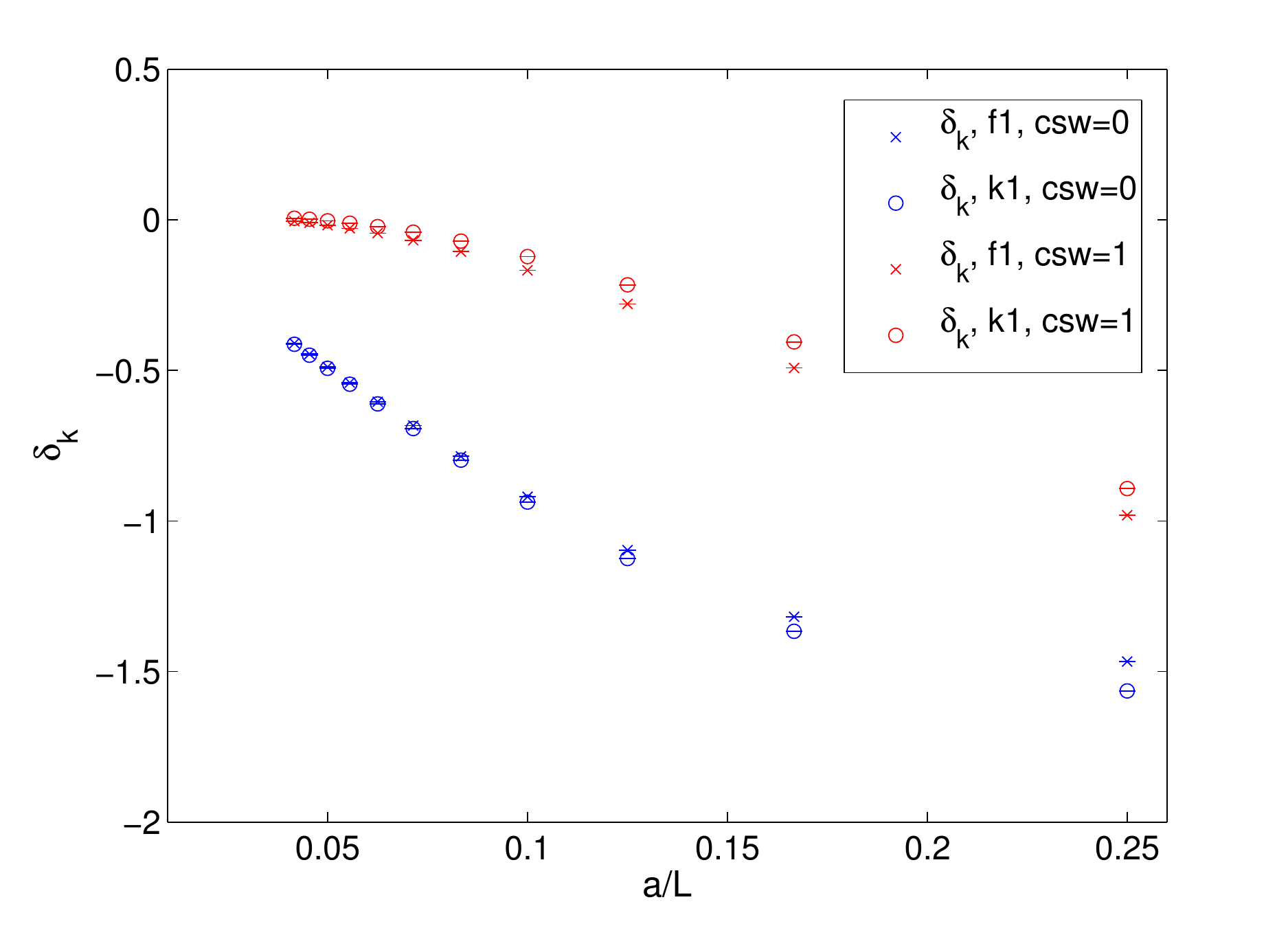}
\label{fig:cutoff_R}
\end{subfigure}
\vspace{-0.05\textwidth}
\caption{1-loop lattice artefacts in the SSFs. On the left, the three lines correspond to the SW action for three values of $\theta$ angle and the two different markers correspond to the two schemes $\alpha=0,1/2$ (note that at 1-loop $f_1=k_1$ for $\theta=0$). On the right, a comparison between cutoff effects from Wilson and SW actions at $\theta=0.5$ . }
\label{fig:cutoff}
\end{figure}

We can expand all the correlators entering Eq. \eqref{eq.renorm}, as well as renormalization constants, in powers of $g_0^2$
\begin{equation}
X=\sum_{n=0}^{\infty}g_0^{2n}X^{(n)},
\label{eq.genericexpansion}
\end{equation}
%\begin{equation}
%Z_T(g_0.L/a)=1+\sum_{n=1}^{\infty}g_0^{2n}Z_T^{(n)}(L/a), \quad k_T(x_0)=\sum_{n=0}^{\infty}g_0^{2n}k_T^{(n)}(x_0)
%\end{equation}
where $X$ in Eq. \eqref{eq.genericexpansion} can be either $Z_T$, $k_T$, $k_V$, $f_1$ and $k_1$. The one-loop improvement for the $k_T(x_0)$ now reads
\begin{equation}
k_T^I(x_0)=k_T^{(0)}(x_0)+g_0^2k_T^{(1)}(x_0)+\left . ag_0^2c_T^{(1)}\tilde{\partial}_0k_V^{(0)}(x) \right |_{x_0} + \mathcal{O}(ag_0^4)
\end{equation}
and the renormalization constant for $\alpha=0$ is given by
\begin{equation}
Z_T^{(1)}(x_0,L/a)=-\left \{ \frac{\bar{k}_T^{(1)}(x_0)}{k_T^{(0)}(x_0)}-\frac{\bar{f}_1^{(1)}}{2f_1^{(0)}}+ac_T^{(1)}\frac{\tilde{\partial}_0k_V^{(0)}(x)|_{x_0}}{k_T^{(0)}(x_0)} \right \}
\end{equation}
where the notation $\bar{F}^{(1)}=F^{(1)}+F^{(1)}_{bi}+m_c^{(1)}\frac{\partial}{\partial m_0}F^{(0)}$ stands for one-loop coefficients where the contribution from boundary counter terms, as well as the one related to the critical mass, have been subtracted \cite{Palombi:2005zd}. The 1-loop critical mass $m_c^{(1)}$ is taken from \cite{Panagopoulos:2001fn}. 
Following \cite{Sint:1998iq} and \cite{Palombi:2005zd}, in order to study the approach to the continuum limit of the SSFs we define the relative deviation 
\begin{equation}
\Delta_k(g_0^2,L/a)=\frac{\Sigma_T(g_0^2,L/a)|_{u=g^2_{SF}(L)}-\sigma_T(u)}{\sigma_T(u)}=\delta_kg_0^2+\mathcal{O}(g_0^4)
\end{equation}
where the one-loop coefficient (see Fig. \eqref{fig:cutoff}) is given by
\begin{equation}
\delta_k=\frac{Z_T^{(1)}(2L/a)-Z_T^{(1)}(L/a)}{\gamma_0\log(2)}-1
\end{equation}
%cutoff
%\vspace{-0.05\textwidth}
%test t <-> H
%\vspace{-0.03\textwidth}
The dependence on $a/L$ of the 1-loop renormalization constant can be described according to \cite{Bode:1999sm} as
\begin{equation}
Z_T^{(1)}(L/a) \approx \sum_{\nu=0}^{\infty}\left (\frac{a}{L} \right )^{\nu} \left \{ r_{\nu} + s_{\nu} \log \left ( \frac{L}{a} \right ) \right \}.
\label{eq.Zfit}
\end{equation}
In order to assess the systematic uncertainties in fit coefficients, we tried several fit ansaetze by truncating the series at different orders and changing the minimal value in the fit range $[(L/a)_{\rm min}$ $, L/a=48]$. We checked that, within systematic uncertainties, the correct value of the LO anomalous dimension $s_0=\gamma_0$ is reproduced; after that, the term with $s_0$ can be subtracted to improve the fitting precision. In the errors on the fit parameters the systematics related to the choice of the ansatz has been taken into account in a conservative way.
In order to extract the two-loops anomalous dimension in the SF scheme for the tensor bilinear, we used a scheme-matching procedure with a given reference scheme where $\gamma_1$ is known. This bypasses a direct two-loop calculation in the SF. Since tensor bilinears renormalize multiplicatively, the same strategy employed for quark masses \cite{Sint:1998iq} can be used. SF NLO anomalous dimension  can be written as 
\begin{equation}
\gamma_{SF}^{(1)}=\gamma_{ref}^{(1)}+2b_0\chi^{(1)}-\gamma_0\chi_g^{(1)}
\label{eq.gamma1}
\end{equation}
where $\chi_g^{(1)}$ is related to a scheme matching for the coupling, and $\chi^{(1)}$ is related to the finite part of Eq. \eqref{eq.Zfit}. In fact   
\begin{equation}
\chi^{(1)}=\chi_{SF,ref}^{(1)}=\chi_{SF,lat}^{(1)}-\chi_{ref,lat}^{(1)} \quad , \quad \chi_g^{(1)}=2b_0\log(\mu L) - \frac{1}{4\pi}(c_{1,0}+c_{1,1}N_f)
\end{equation}
and $\chi_{SF,lat}^{(1)}=r_0$. Since Eq. \eqref{eq.gamma1} is independent on the choice of the ``ref'' scheme, we have crosschecked our results using both $\overline{\rm MS}$ and $\rm RI'$ schemes (\cite{Skouroupathis:2008mf}, \cite{Capitani:2000xi} for finite parts, and \cite{Gracey:2003mr} for the two-loops anomalous dimension in both schemes).
%Z 1-loop
%\vspace{-0.025\textwidth}
\begin{figure}[t]
\begin{subfigure}{0.51\linewidth}
\includegraphics[width=\linewidth]{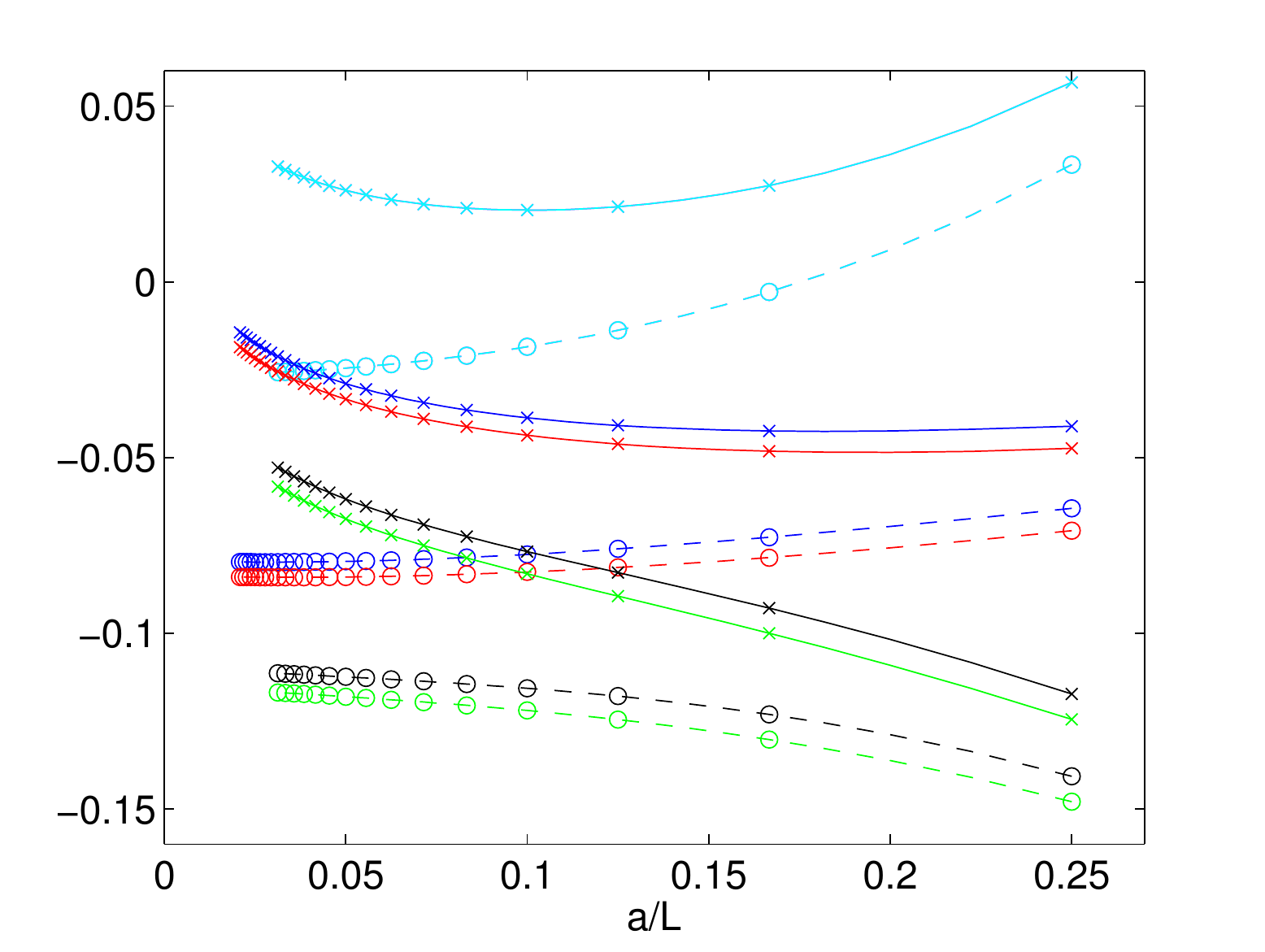}
\label{fig:Z1loop_L}
\end{subfigure} \hspace{0.015\linewidth}
\begin{subfigure}{0.26\linewidth}
\includegraphics[width=\linewidth]{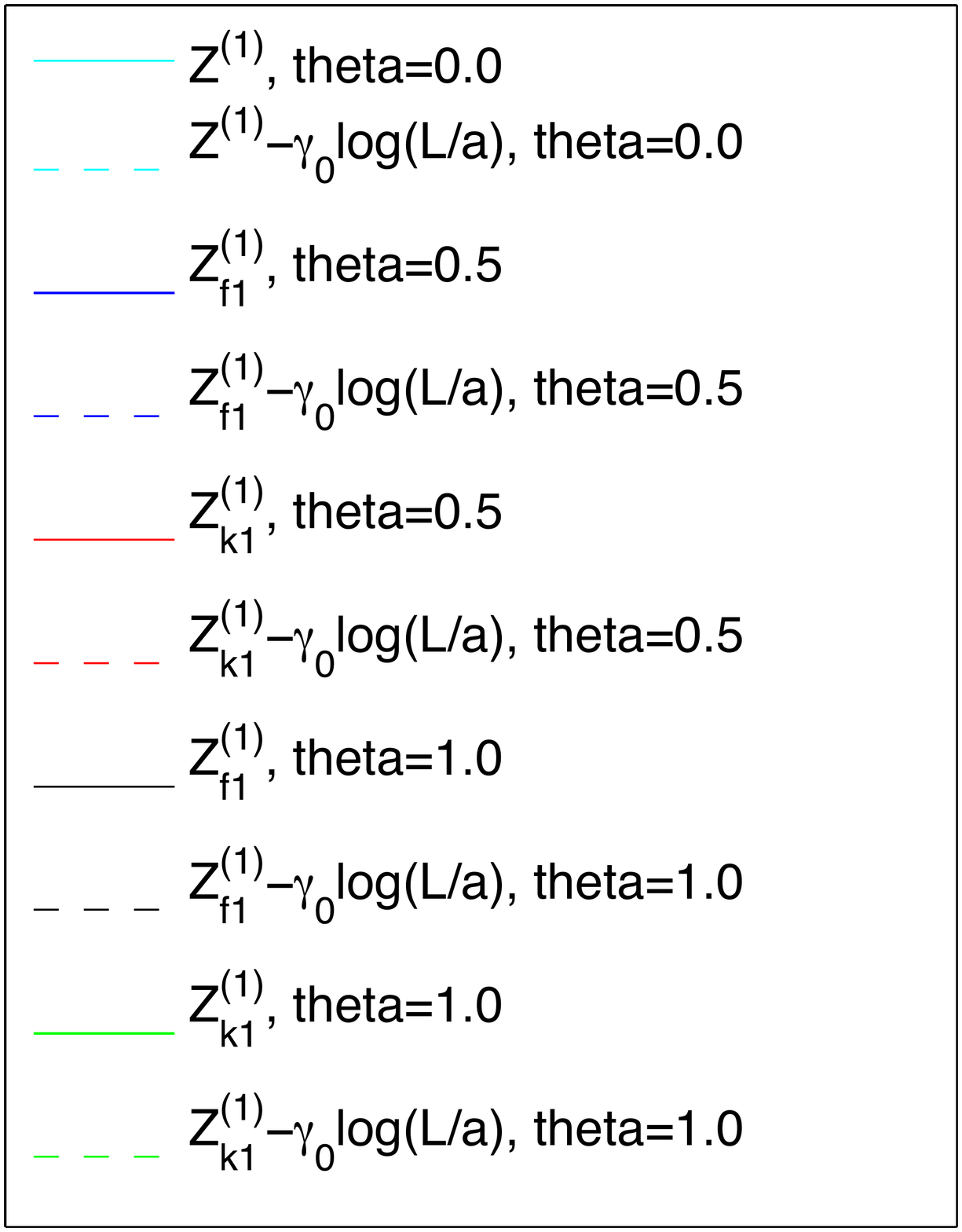}
\label{fig:legendZ1loop_R}
\end{subfigure}
\vspace{-0.05\textwidth}
\caption{$Z_T^{(1)}$ at 1-loop for various values of $\theta$ and for $\alpha$ (in particular the two choices of $\alpha$ correspond to scheme labeled $f_1$ and $k_1$). The behaviour of the renormalization constant after the subtraction of the leading logaritmic divergence is also displayed. }
\label{fig:Z1loop}
\end{figure}
%table gamma1/gamma0
\begin{table}[b]
  \begin{center}
    \label{tab:gamma1_o_gamma0}
    \begin{tabular}{|c|c|c|c|}
    \hline
     scheme s &  SF ($\alpha=0$) & SF ($\alpha=1/2$) & $\overline{\rm MS}$ \\
      \hline
       $\gamma_1^{\rm s}/\gamma_0$ & $0.4113(33)-0.0132(1)N_f$ & $0.3767(33)-0.0111(1)N_f$ & $0.1910-0.091 N_f$\\
       \hline
    \end{tabular}
        \caption{NLO anomalous dimension in the SF scheme for the two schemes corresponding to $\alpha=0,1/2$ and $\theta=0.5$, compared with the anomalous dimensions in continuum schemes.}
  \end{center}
\end{table}
\vspace{-0.02\textwidth}
\section{Non-perturbative renormalization and running}
Our non-perturbative computation has been carried out for both $N_f=0,2$ and is ongoing for $N_f=3$ in parallel with the mass \cite{Campos:2015fka}. Once the renormalization constants given by imposing the condition \eqref{eq.renorm} have been computed on a given lattice of size $L/a$ and the double lattice of size $2L/a$,  eq. \eqref{eq.ssf} is used to obtain the non-perturbative value of the SSF. We have computed here the SSFs for $14$ values for the coupling in the range $u=[0.8873,3.480]$ for quenched data, and for $6$ couplings in the range $u=[0.9793,3.3340]$ for $N_f=2$. Since we did not implement $\mathcal{O}(a)$ improvement of the tensor operator in the quenched case, the continuum extrapolation is linear in $a/L$ and has been performed using lattices with $L/a=\{6,8,12,16\}$ and $2L/a=\{12,16,24,32\}$. In the final analysis the smallest lattice has been discarded because it is affected by large cutoff effects. In order to reduce the uncertainty on the extrapolation we have performed a constrained fit between data from an $\mathcal{O}(a)$-improved action and an unimproved one after testing universality of the continuum limit. Regarding $N_f=2$, since both action and operator are $\mathcal{O}(a)$-improved, the continuum limit is approached quadratically. In this case the extrapolation has been performed using $L/a=\{6,8,12\}$ and $2L/a=\{12,16,24\}$. Once in the continuum we adopted a polynomial fit ansatz for the SSFs of the form
\begin{equation}
\sigma_T(u)=1+\sigma^{(1)}u+\sigma^{(2)}u^2+\sigma^{(3)}u^3 + \mathcal{O}(u^4)
\end{equation}
where the first two coefficients are kept fixed to their perturbative values
\begin{equation}
\sigma^{(1)}=\gamma_0\log(2) \quad , \quad \sigma^{(2)}=\gamma_1^{\rm SF}\log(2)+ \left [ \frac{1}{2} (\gamma_0)^2+b_0\gamma_0 \right ] (\log(2))^2
\end{equation}
and the cubic coefficient has been fitted for both quenched and two-flavour data. 
%\vspace{-0.02\textwidth}
\begin{figure}[t]
\begin{subfigure}{0.33\linewidth}
\includegraphics[width=\linewidth]{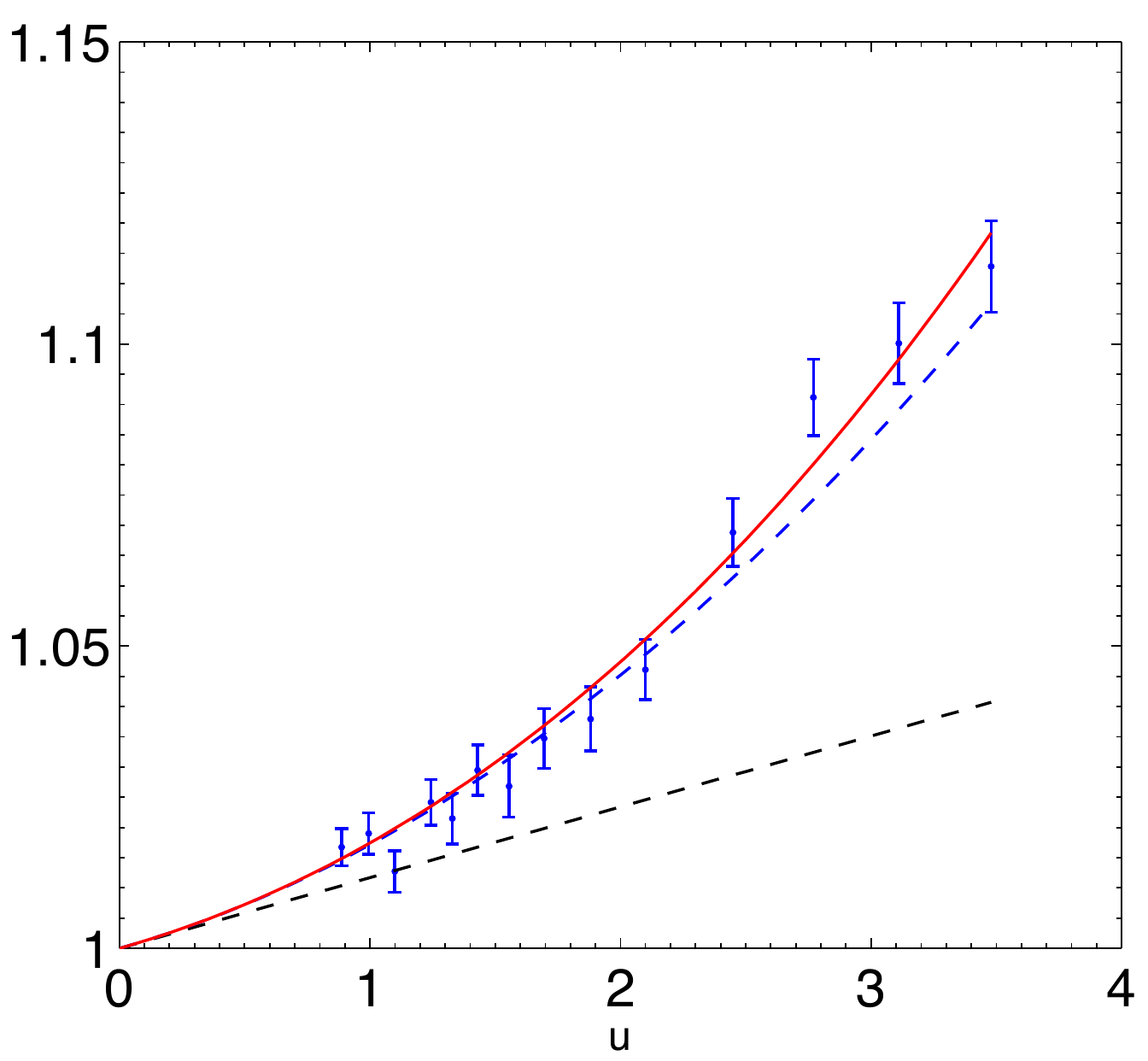}
\label{fig:SSF_nf0_L}
\end{subfigure} \hspace{-0.0125\linewidth}
\begin{subfigure}{0.33\linewidth}
\includegraphics[width=\linewidth]{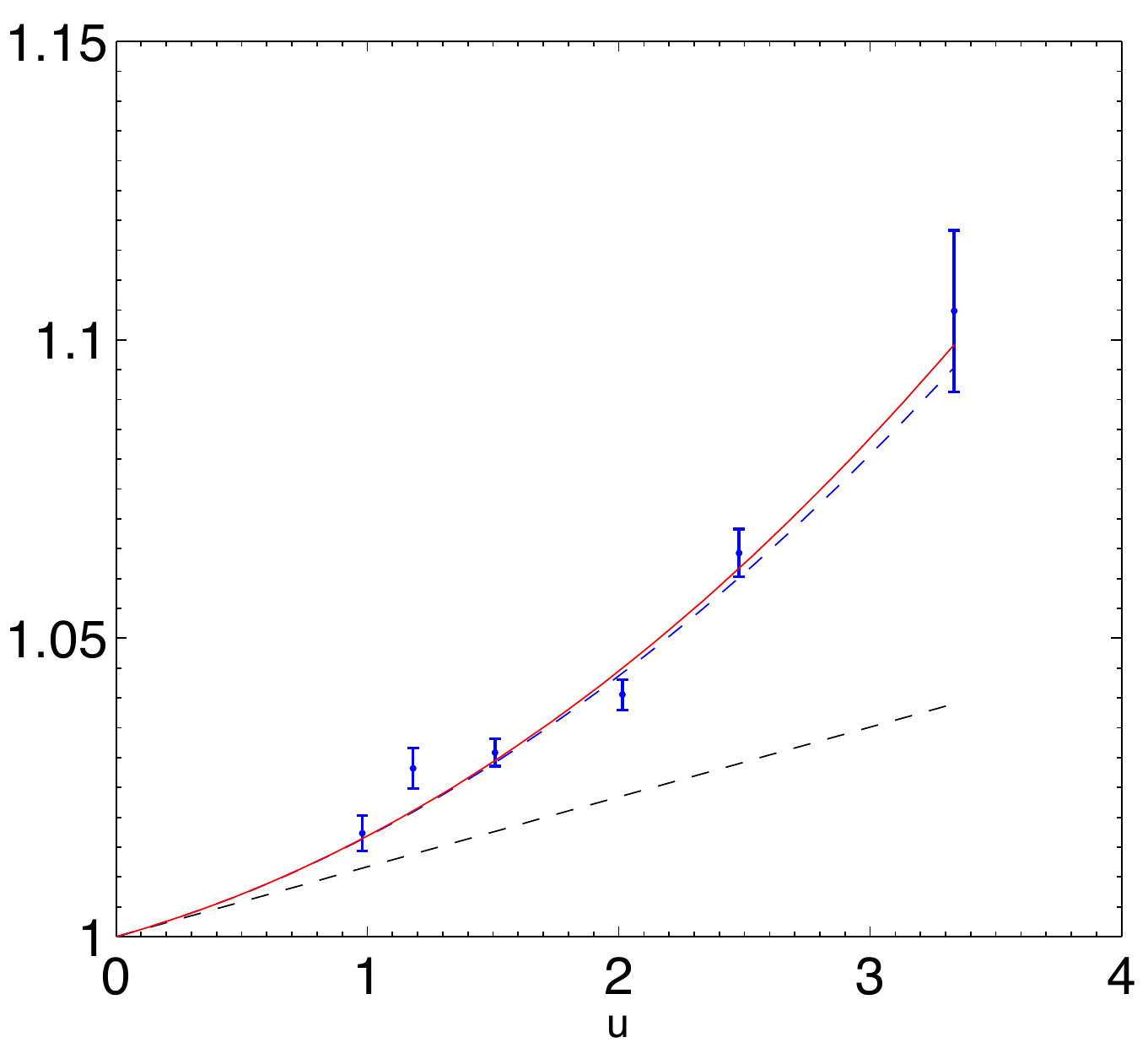}
\label{fig:SSF_nf2_R}
\end{subfigure}\hspace{-0.005\linewidth}
\begin{subfigure}{0.319\linewidth}
\includegraphics[width=\linewidth]{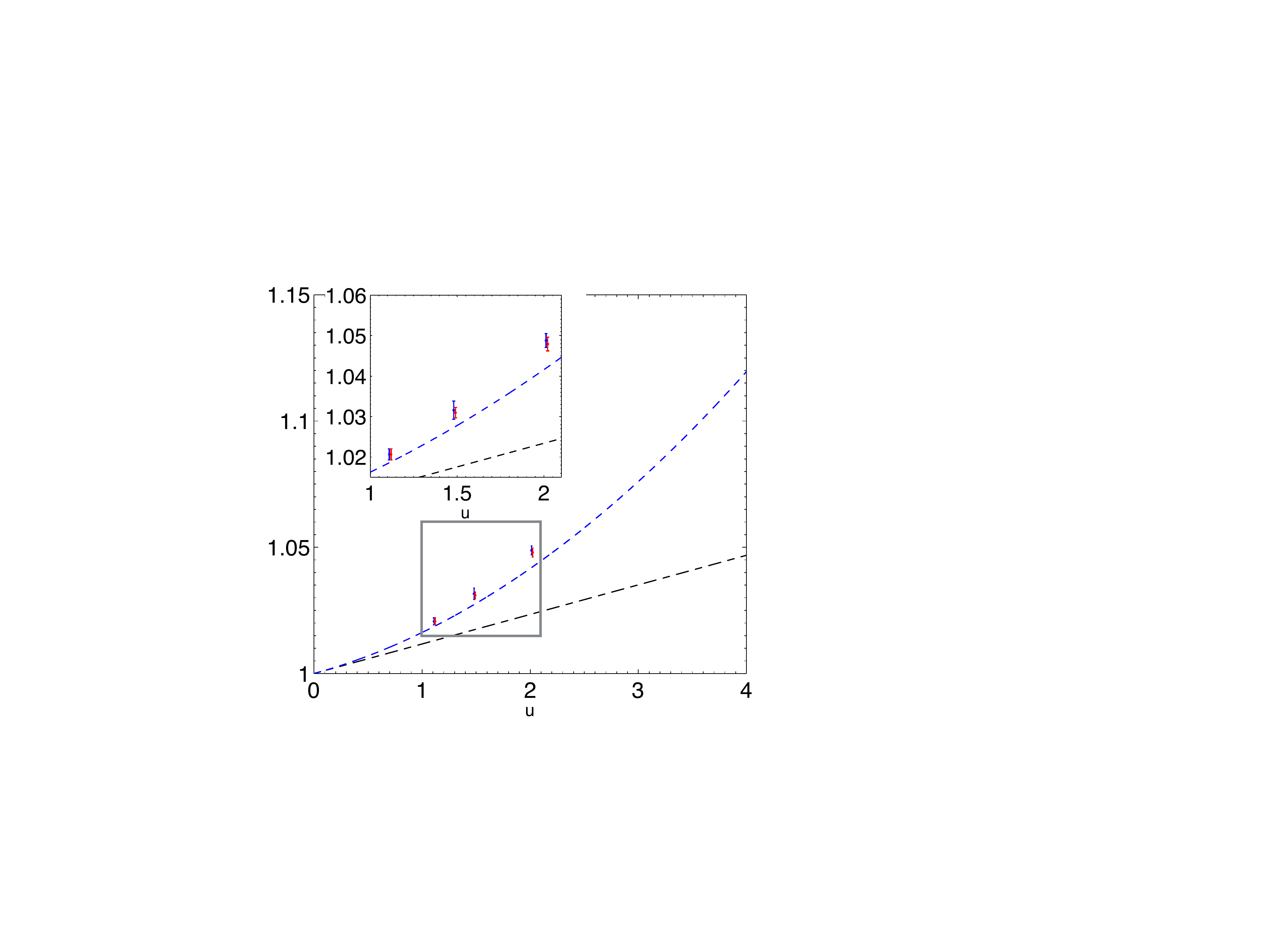}
\label{fig:SSF_nf2_R}
\end{subfigure}
\label{fig:SSF_nf02}
\vspace{-0.04\textwidth}
\caption{SSFs for $N_f=0$ (left figure), $N_f=2$ (center figure), preliminary $N_f=3$ (on the right). Dashed black line is the LO approximation, dashed blue line is the NLO approximation, red solid line is the fit including the cubic term. Blue markers are non-perturbative points. In the plot on the right red and blue points correspond to unimproved and 1-loop improved SSF respectively.}
\end{figure}
%\vspace{-0.0304\textwidth}
The running between two scale $\mu_1$ and $\mu_2$ is given by eq. \eqref{eq.CS} and can be written as
\begin{equation}
U(\mu_2,\mu_1)=\exp \left \{ \int_{\bar{g}(\mu_1)}^{\bar{g}(\mu_2)} \, dg \, \frac{\gamma(g)}{\beta(g)} \right \}=\lim_{a \to 0} \, \frac{Z(g_0,a\mu_2)}{Z(g_0,a\mu_1)}. 
\label{eq.run1}
\end{equation}
Once SSF has been fitted on the range of couplings, the non-perturbative running can be obtained. The evolution coefficient is computed non-perturbatively as a product of SSFs $U(\mu_{pt},\mu_{had})=\prod_{i=1}^{n}[\sigma_T(u_i)]^{-1}$ with $u_i=\bar{g}^2(2^i\mu_{had})$. For both $N_f=0$ and $N_f=2$ we have computed $n=7$ non-perturbative steps (i.e. achieving a factor $2^7$ in the ratio between $\mu_{pt}$ and $\mu_{had}$), connecting an hadronic scale $\mu_{had}=275 \, \rm MeV$ ($476 \, \rm MeV$ for $N_f=2$) up to $\approx35\, \rm GeV$ ($\approx 62 \, \rm GeV$) an high energy scale, where perturbation theory is supposed to be safe. At those scales (computed with $\Lambda_{\rm SF}$ for $N_f=0$ and $N_f=2$ from \cite{Luscher:1993gh,Fritzsch:2012wq} respectively) the NP evolution is matched with perturbation theory at NLO, $\hat{c}(\mu_{had})=\hat{c}(\mu_{pt})U(\mu_{pt},\mu_{had})$ where $\hat{c}$ is defined as
\begin{equation}
\hat{c}(\mu)=\frac{O_{RGI}}{O(\mu)}=\frac{Z_T^{\rm RGI}}{Z_T(\mu)}=\left [ \frac{\bar{g}^2(\mu)}{4\pi} \right ]^{-\frac{\gamma_0}{2b_0}}\exp \left \{ -\int_0^{\bar{g}(\mu)} \, dg \, \left ( \frac{\gamma(g)}{\beta(g)} -\frac{\gamma_0}{b_0g} \right ) \right \}
\end{equation}
Equivalently, the total RGI renormalization constant is defined as \begin{equation}Z_T^{\rm RGI}(g_0)=\hat{c}(\mu_{pt})U(\mu_{pt},\mu_{had})Z_T(g_0,\mu_{had}).\end{equation} 
\vspace{-0.03\textwidth}
\begin{figure}[h]
\begin{subfigure}{0.48\linewidth}
\includegraphics[width=\linewidth]{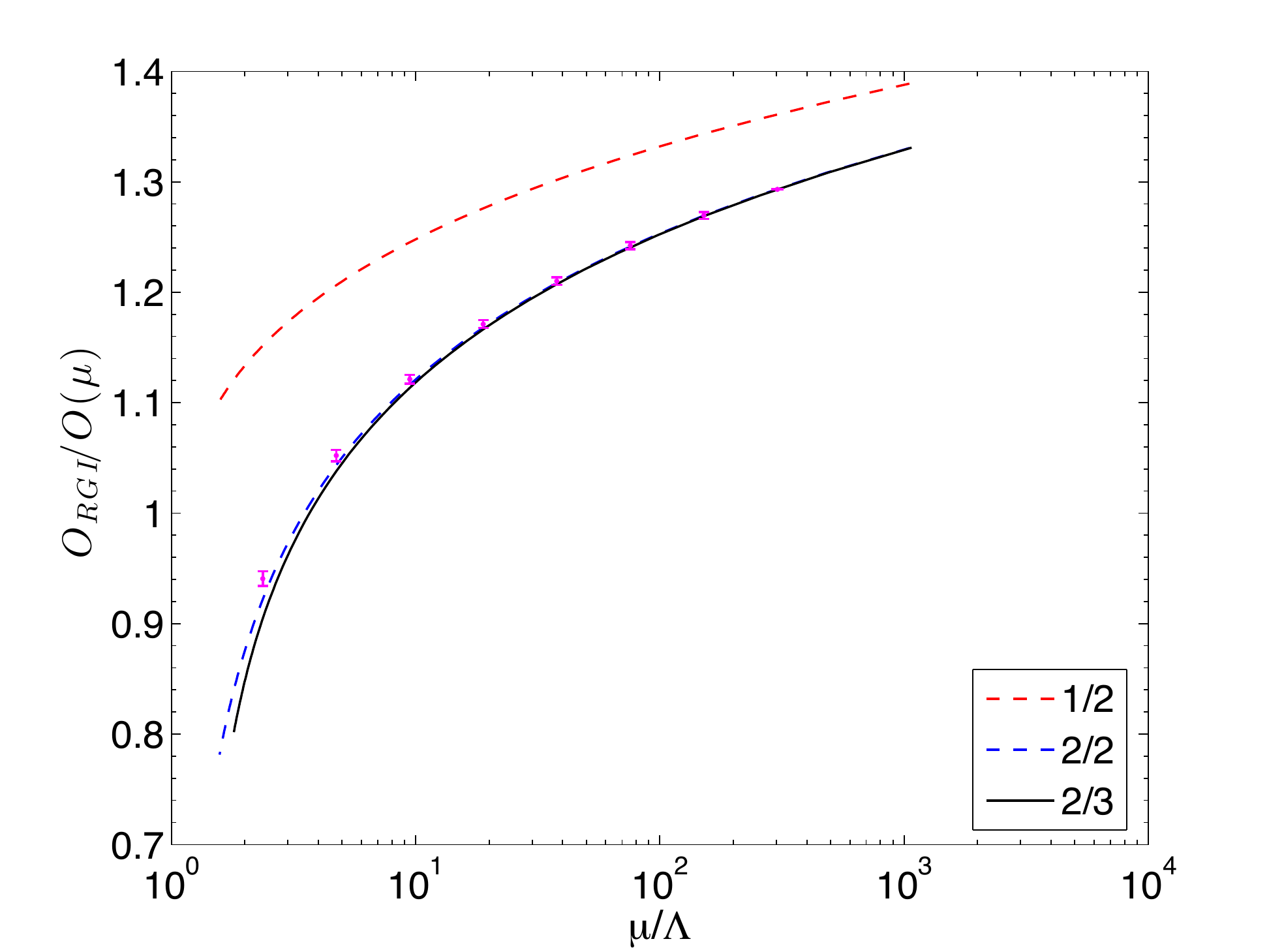}
\label{fig:SSF_nf0_L}
\end{subfigure} \hspace{0.01\linewidth}
\begin{subfigure}{0.48\linewidth}
\includegraphics[width=\linewidth]{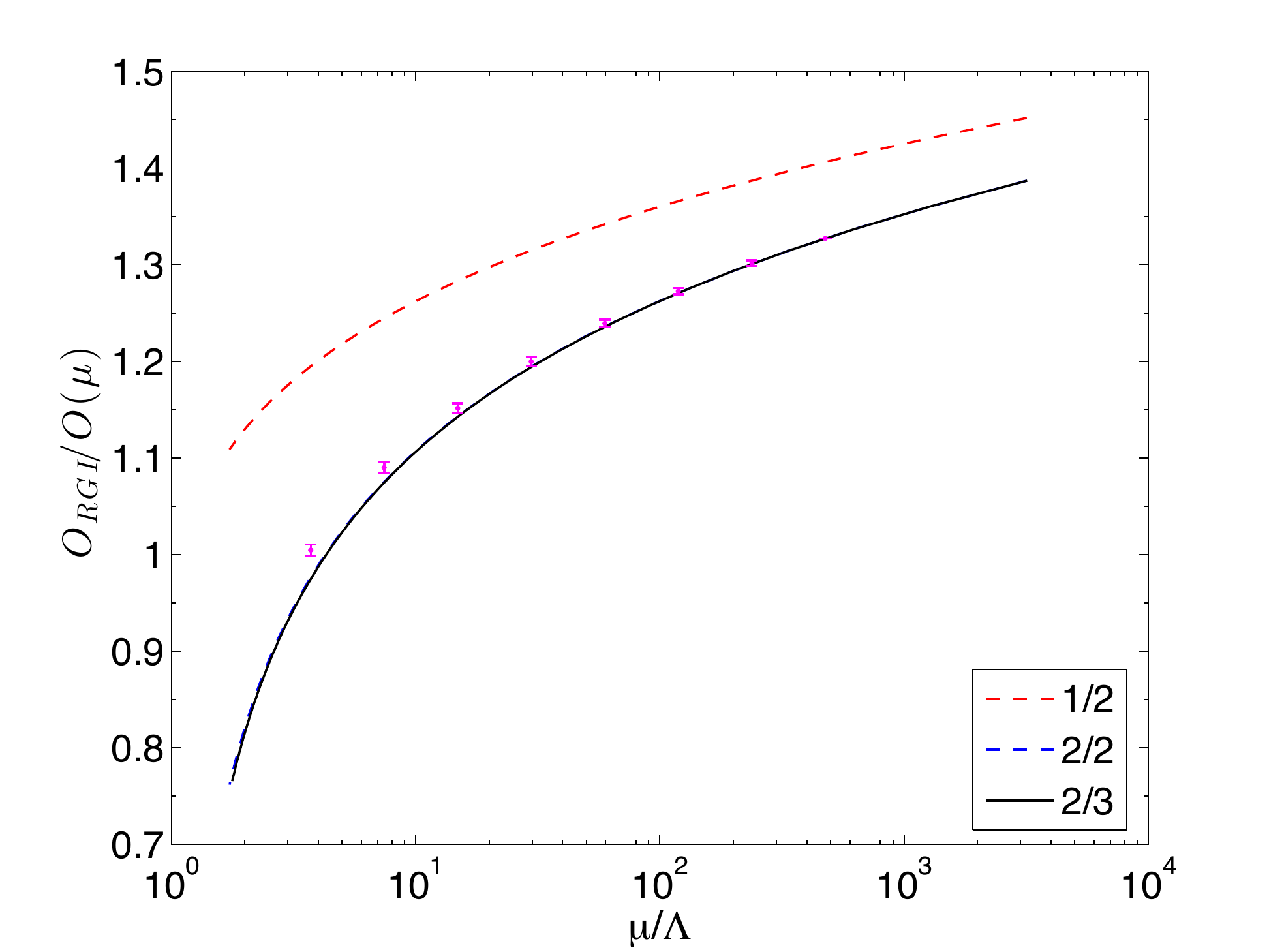}
\label{fig:SSF_nf2_R}
\end{subfigure}
\label{fig:SSF_nf02}
\vspace{-0.04\textwidth}
\caption{The red dashed line is the LO running coefficient $\hat{c}$ while the dashed blue line and solid black line are NLO approximation the first with NLO-$\gamma$ and NLO-$\beta$, the latter with NLO-$\gamma$ and NNLO-$\beta$. Results for $N_f=0$ (on the left) and $N_f=2$ (on the right) are provided in a single scheme ($\alpha=0$).}
\end{figure}
\vspace{-0.06\textwidth}
\section{Conclusions}
On the perturbative side we have analysed cutoff effects of the SSF for various SF schemes, and from the finite parte of the 1-loop renormalization constant we have been able to give the first preliminary determination of the NLO anomalous dimension in SF schemes for the tensor currents, which are the only quark bilinears with a non-trivial anomalous dimension independent from that of quark masses. Moreover, thanks to the non-perturbative lattice computation for both $N_f=0$ and $N_f=2$ we have computed the non-perturbative SSF in the continuum through which with 7 recursion steps the running over more than 2 orders of magnitude is computed, from an hadronic scale up to a perturbative one. Despite the dependence of the running on the scheme and on $N_f$, since the correction given to the running by the NLO anomalous dimension respect to the LO is large, we still observe large systematics due to the matching with perturbation theory on the scale of $2$-$3 \, \rm GeV$. The same strategy, briefly explained here, is being applied to $N_f=3$ simulations that will allow a more physical and precise determination of both the running and the RGI.

\vspace{-0.02\textwidth}

\end{document}